\documentclass[preprint,aps,amsmath,superscriptaddress,nofootinbib
]{revtex4}
\usepackage{dcolumn}
\usepackage{bm}
\usepackage{epsfig}
\usepackage{hyperref}
\usepackage{graphicx}
\usepackage{dcolumn}
\usepackage{bm}
\usepackage{longtable}
\begin{document}

\title{Couplings of the Rho Meson in a Holographic dual of QCD with Regge Trajectories}
\author{Tao Huang} \thanks{Email: huangtao@mail.ihep.ac.cn.}
\affiliation{Institute of High Energy Physics, Chinese Academy of
Sciences, Beijing 100049, China}

\author{Fen Zuo} \thanks{Email: zuof@mail.ihep.ac.cn.}
\affiliation{Department of Modern Physics, University of Science and
Technology of China, Hefei, Anhui 230026, China}

\affiliation{Institute of High Energy Physics, Chinese Academy of
Sciences, Beijing 100049, China}

\begin{abstract}
The couplings $g_{\rho HH}$ of the $\rho$ meson with any hadron
$H$ are calculated in a holographic dual of QCD where the Regge
trajectories for mesons are manifest. The resulting couplings grow
linearly with the exciting number of $H$, thus are far from
universal. A simple argument has been given for this behavior
based on quasi-classical picture of excited hadrons. It seems that
in holographic duals with exact Regge trajectories the $g_{\rho
HH}$ universality should be violated. The $\rho$-dominance for the
electromagnetic form factors of $H$ are also strongly violated,
except for the lowest state, the pion. Quite unexpected, the form
factor of the pion is completely saturated by the contribution of
the $\rho$. The asymptotic behavior of the form factors are also
calculated, and are found to be perfectly accordant with the
prediction of conformal symmetry and pertubative QCD.
\end{abstract}



\maketitle

\section{Introduction}
~~In 1998, Maldacena advocated the famous duality between string
theory on Anti-de Sitter (AdS) space and certain conformal field
theory (CFT) on the boundary, now known as AdS/CFT correspondence
 \cite{maldacena}. Based on this duality, it was shown by Polchinski
that the power law behavior of the high-energy fixed angle
scattering of glueballs in confining gauge theories could be derived
nonperturbatively in the dual theory \cite{polchinski}, and was
found to be precisely as the simple dimensional prediction in QCD
\cite{brodsky}. Since then, various pertubations have been
introduced to the original AdS background to produce supergravity
duals with a mass gap, confinement, and chiral breaking, with the
purpose of establishing a exact dual of the true QCD. One can
introduce a simple cutoff in the radical coordinate to simulate
color confining, and get the so called the hard-wall model
\cite{polchinski}. With only one parameter for the cutoff, the lower
excited hadron spectrum in this model are found to be perfectly
agree with the physical states \cite{brodsky1}. Moreover, chiral
breaking can be modeled rather well in this model too, with two more
parameters: the current quark mass and the quark condensate
\cite{katz1}. One can also introduce fundamental quarks by adding
$D7$ branes to the original background, resulting the $D3/D7$ model,
in which the chiral breaking can be incorporated rather naturally
\cite{katz2}. However, in both models, the masses of highly excited
mesons grow linearly with the exciting number, which conflict with
the familiar Regge trajectory \cite{shifman}, where the mass square
grows linearly with the exciting number.

  The couplings $g_{\rho HH}$ of the $\rho$ meson to any hadron $H$
are shown to be quasi-universal in these AdS/QCD models, even when
the $\rho$-dominance is violated \cite{rho}. Actually, as argued
in Ref. \cite{shifman}, this is an "accidental" universality,
since the contributions of excited vector mesons are of the same
order as the rho meson do, but they are sign-alternating and
compensate each other, resulting an apparent $\rho$-dominance.
However, since the meson spectrum obtained in these models has a
wrong dependance on the exciting number, it was argued in Ref.
\cite{shifman} that the $g_{\rho HH}$ universality might be
implement in a wrong way.

The asymptotically linear Regge trajectories for the mesons can be
obtained in the dual theory by adding a non-constant dilaton field
$\Phi$ to the original  AdS$_5$ background \cite{dilaton}. The
Though \textit{ad hoc}, the special profile for the dilaton seems to
be necessary to guarantee the linear Regge behavior \cite{gursoy}.
One can also introduce a gaussian warp factor to the AdS$_5$ metric,
which is shown to be equivalent to the previous background when
studying the meson spectrum \cite{Andreev}.

In this paper we would like to discuss if the $g_{\rho HH}$
universality holds in this modified model, which is often referred
as soft-wall model or harmonic oscillator model, or more generally
in AdS/QCD models with linear Regge trajectories. We calculate the
couplings of $g_{\rho HH}$ in this modified model and the results
show that the couplings actually grow linearly with the exciting
number of $H$, thus far from been universal. We can give a simple
explanation for this dependance based on the quasi-classical picture
of excited hadrons. Thus it seems that the $g_{\rho HH}$
universality in Ref.\cite{rho} is somehow connected to the wrong
dependance of the meson mass on the exciting number, and it does not
hold any more when exact Regge behavior is obtained. The
$\rho$-dominance for the electromagnetic form factors of $H$ are
also strongly violated, except for the lowest state, the pion. Quite
unexpected, the form factor of the pion is completely saturated by
the contribution of the $\rho$ meson. The asymptotic behavior of the
form factors are also calculated, and are found to be perfectly
accordant with the prediction of conformal symmetry and pertubative
QCD.

The paper is organized as follows. In the next section we give a
brief review of the soft-wall model and derive the string modes for
the scalar hadron states. In sec.{I}{I}{I} we give explicit results
of $g_{\rho HH}$ and show that the $g_{\rho HH}$ universality
doesn't hold any more in this model. Then we go on to discuss
$\rho$-dominance and asymptotic behavior of the form factors in
Sec.{I}{V}  and {V}, and give a simple summary in the last section.

\section{A brief review of soft-wall model and string modes for scalars}
~~We are in position to work in the background proposed in
\cite{dilaton}. That is, the AdS metric is given by
\begin{equation}
ds^2=e^{2A(z)}(dz^2+\eta_{\mu\nu}dx^\mu~dx^\nu)
\end{equation}
with $A(z)=-\log z$ and turn on a non-constant dilaton field given
by $\Phi(z)=z^2$. We are going to calculate the $\rho$'s couplings
$g_{\rho HH }$ to the scalar states $H$ as a function of the
excitation level $a$ and the twist $\tau$ of the corresponding
operator, as in Ref.\cite{rho}. We have chosen to consider the twist
but not the naive dimension due to the arguments in Ref.
\cite{brodsky2}. First, we have to solve the equation of motion for
the scalar field $s_H(z)$ dual to $H$. Actually this has been done
in discussing the spectrum of the scalar glueball \cite{colangelo}.
It is found that after applying a Bogoliubov transformation
$\psi_H(z)=e^{-B(z)/2}s_H(z)$ with $B(z)=\Phi(z)-3A(z)$, the
function $\psi_H(z)$ satisfies one dimensional Schr\"{o}dinger
equation:
\begin{equation}
-\psi''_H(z)+V_H(z)\psi_H(z)=-q^2\psi_H(z).
\end{equation}
For the normalizable mode $q^2=-m^2$ gives the mass of the
corresponding hadron state. The potential $V_H(z)$ is given by
\begin{equation}
V_H(z)=z^2+15/4z^2+2+m_5^2/z^2,
\end{equation}
with $m_5^2=\tau(\tau-4)$ the AdS mass of the dual field. Then one
can easily get the eigenfunctions
\begin{equation}
\psi^\tau_a(z)=e^{-z^2/2}z^{\tau-3/2}\sqrt{\frac{2a!}{(\tau+a-2)!}}L^{\tau-2}_a(z^2),
\end{equation}
where $L^m_n$ are associated Laguerre polynomials, and the
corresponding eigenvalues are $m^2=4a+2\tau$. After making the
Bogoliubov transformation we get the original mode function:
\begin{equation}
s^\tau_a(z)=z^\tau\sqrt{\frac{2a!}{(\tau+a-2)!}}L^{\tau-2}_a(z^2).
\end{equation}
Certainly $s^\tau_a(z)\to z^\tau$ as $z\to 0$ as it should be.
Notice that if supposing that the variation of $\tau$ is due to
the excitation of the orbital momentum as in Ref.\cite{brodsky1}
$\tau=2+L$, one can get the linear dependance of the meson mass
square on the orbital momentum
\begin{equation}
m^2=4a+2L+4
\end{equation}
though the slopes for the radical number and the orbital momentum
are different. As in Ref.\cite{brodsky1}, we can identify the
$L=0$ states with pion and its radical exciting states. Since the
chiral breaking has not been included, their masses are degenerate
with the corresponding vector mesons \cite{dilaton}.

On the other hand, the solutions for the $\rho$ mesons have been
derived in Ref.\cite{dilaton}, and are given by
\begin{equation}
v_n(z)=z^2\sqrt{\frac{2n!}{(1+n)!}}L^1_n(z^2)
\end{equation}
with squared masses of the $\rho$s $m_n^2=4(n+1)$, and the decay
constants:
\begin{equation}
F_{\rho_n}^2=\frac{1}{g_5^2}[v_n''(0)]^2=\frac{8(n+1)}{g_5^2}
\end{equation}
where $g_5$ is the 5D coupling of the dual vector field for rho.
Here $n$ is supposed to be the radical number and $n=0$ gives the
$\rho$ meson.

\section{the $\rho$ Couplings $g_{\rho HH}$}
~~Now we can calculate the couplings $g_{\rho HH}$ using the
formula \cite{rho1}:
\begin{equation}
g_{\rho HH}=g_5\int{\mu_H(z)v_0(z)s_H(z)^2} \label{eq:coupling1}
\end{equation}
with $\mu_H(z)$ in the above background given by
\begin{equation}
\mu_H(z)=e^{-B(z)}=e^{-\Phi(z)+3A(z)}.
\end{equation}
The resulting couplings are:
\begin{equation}
F_{\rho}g_{\rho aa}^{(2)}/m_{\rho}^2=1,~3,~5,~7,~9,...
\end{equation}
for $a=0,~1,~2,~3,~4$; and
\begin{equation}
F_{\rho}g_{\rho 00}^{(\tau)}/m_{\rho}^2=1,~2,~3,~4,~5,...
\end{equation}
for $\tau=2,~3,~4,~5,~6$. It seems that $g_{\rho HH}$ actually
grows linearly with the exciting number $a$ and the twist
dimension $\tau$ of $H$. This is quite different from the behavior
of $g_{\rho HH}$ in the hard-wall and the D3/D7 model considered
in Ref.~\cite{rho}, where $g_{\rho HH}$ are shown to be
quasi-universal and lie within a narrow band near
$m_\rho^2/F_\rho$. Notice that in both the models there the meson
masses grow linearly with the radical number, so the $g_{\rho HH}$
universality in these models seems to be connected to this wrong
dependance. Our results show that when the dependance is
corrected, the $g_{\rho HH}$ universality doesn't hold any more.

To make this more clear, let's analyze the integral in
Eq.(\ref{eq:coupling1}) carefully. As shown in Ref.\cite{rho}, the
$g_{\rho HH}$ universality in generic AdS/QCD models is based on two
facts. First, there is a maximal value $z=z_{\rm{max}}$ beyond which
scale-invariance is badly broken; second, being the lowest mode of a
conserved current, the $\rho$ meson should be structureless and have
no nodes, and must satisfy Neumann boundary conditions at
$z=z_{\rm{max}}$. In other words, $v_0(z)$ will always increase
monotonously with $z$ and reach it's maximal value at
$z=z_{\rm{max}}$. Thus the integral in Eq.(\ref{eq:coupling1}) in
always dominated by the contributions in the region $z\sim
z_{\rm{max}}$ where $v_0(z)$ varies slowly and has a typical value
$\hat v_0(z)$. Replacing $v_0(z)$ by $\hat v_0(z)$ in
Eq.(\ref{eq:coupling1}) and using the normalization condition for
$s_H(z)$, we finally get the universal coupling $g_{\rho HH}$ and
can actually prove that it equals to the expected value
$m_\rho^2/f_\rho$ from $\rho$-dominance \cite{rho}. However, in the
soft-wall model, there isn't an absolute cutoff $z_{\rm{max}}$.
Based on the quasi-classical arguments in Ref.\cite{shifman}, one
can expect that this cutoff should be proportional to the length
$L_n=M_n/\sigma$ of the flux tube of the excited meson, where
$\sigma$ is the tension of the flux tube. Note $v_0(z)\sim z^2$ and
the linear relation of mass square $m_n^2\sim n$, one should then
replace $v_0(z)$ by $z^2\sim L_n^2\sim n$. Thus one gets a linearly
increasing coupling $g_{\rho HH}$.

\section{violation of the vector meson dominance} ~~It is well known
that, the $\rho$ meson coupling universality can be induced from the
VMD hypothesis,  that is the $\rho$ meson gives the dominant
contribution to the electromagnetic form factor of the various
hadrons. Now since the $g_{\rho HH}$ universality doesn't hold any
more, we can conclude that the $\rho$ meson dominance must be
violated at the same time. Then we go ahead to calculate the
couplings of the excited $\rho$ mesons with $H$. The calculation is
straight forward, we just need to replace the $\rho$ mode in
Eq.(\ref{eq:coupling1}) by the corresponding exciting $\rho$ states.
For the radical exciting states, the results are as follows: for
$n=0,~1,~2,~3,~4,...$
\begin{eqnarray}
F_{\rho_n}g_{\rho_n 00}^{(2)}/m_{\rho_n}^2&=&1,~0,~0,...\\\nonumber
F_{\rho_n}g_{\rho_n
11}^{(2)}/m_{\rho_n}^2&=&3,-4,~2,~0,~0...\\\nonumber
F_{\rho_n}g_{\rho_n
22}^{(2)}/m_{\rho_n}^2&=&5,-14,~22,-18,~6,~0,~0,...
\end{eqnarray}
 For the orbital exciting states, we have:
\begin{eqnarray}
F_{\rho_n}g_{\rho_n 00}^{(2)}/m_{\rho_n}^2&=&1,~0,~0,...\\\nonumber
F_{\rho_n}g_{\rho_n
00}^{(3)}/m_{\rho_n}^2&=&2,-1,~0,~0,...\\\nonumber
F_{\rho_n}g_{\rho_n 00}^{(4)}/m_{\rho_n}^2&=&3,-3,~1,~0,~0,...
\end{eqnarray}
with all the neglected terms vanishing. From the result follows
that, the electromagnetic form factor of the pion is completely
saturated by the contribution of the $\rho$ meson, given exactly
the phenomenologically successful VMD fit
\begin{equation}
F_\pi^{\rm{VMD}}(Q^2)=1/(1+Q^2/m_\rho^2)
\end{equation}
The VMD in the electromagnetic form factor of the pion was also
found in the hard-wall model in Ref.\cite{pomarol} in considering
chiral symmetry breaking. Except for the pion, there is no evidence
of VMD in the form factors of the other hadrons. Actually, the
contribution of some exciting state may exceed that of $\rho$ by
several times. The vector meson dominance is certainly violated very
badly, just as what we have predicted from the previous calculation.
However, the contributions of various $\rho$ mesons indeed show a
sign-alternating feature, which is supported by QCD analysis
\cite{shifman}.

\section{the asymptotic behavior of the electromagnetic form factor}
~~Another issue we want to emphasize is the asymptotic behavior of
the form factor. This can be easily obtained from the previous
results and the decomposition formula \cite{rho1}:
\begin{equation}
F_H(q^2)=\sum_n{\frac{F_{\rho_n}g_{{\rho_n}HH}}{q^2+m_{\rho_n}^2}}
\end{equation}
We find, for the hadrons created by the lowest twist operators,
all the radical states exhibit the same asymptotic behavior
exactly:
\begin{equation}
F^{(\tau=2)}_{aa}(q^2)\to (\tau-1)m_\rho^2/q^2.
\end{equation}
This is obvious from the VMD for the pion, but not so straight for
the other radical states. The asymptotic behavior of the pion has
been analyzed utilizing the Light-Cone wave function dual to the
string modes in this model, and the same results was obtained in
Ref.\cite{brodsky3} and further in Ref.\cite{brodsky4}. For the
higher twist states, the results are radical number dependant, and
we give the results for the first few states only:
\begin{eqnarray}
F^{(\tau=3)}_{00}(q^2)&\to&(\tau-1)\frac{
m_\rho^2}{q^2}\left(\frac{m_\rho^2}{q^2}\right)\\\nonumber
F^{(\tau=3)}_{11}(q^2)&\to&(\tau-1)\frac{
m_\rho^2}{q^2}\left(\frac{m_{\rho_1}^2}{q^2}\right)\\\nonumber
F^{(\tau=3)}_{22}(q^2)&\to&(\tau-1)\frac{
m_\rho^2}{q^2}\left(\frac{m_{\rho_2}^2}{q^2}\right)\\\nonumber
F^{(\tau=4)}_{00}(q^2)&\to&(\tau-1)\frac{
m_\rho^2}{q^2}\left(\frac{m_\rho^2}{q^2}\frac{m_{\rho_1}^2}{q^2}\right)\\\nonumber
F^{(\tau=4)}_{11}(q^2)&\to&(\tau-1)\frac{
m_\rho^2}{q^2}\left(\frac{m_{\rho_1}^2}{q^2}\frac{m_{\rho_2}^2}{q^2}\right)\\\nonumber
F^{(\tau=4)}_{22}(q^2)&\to&(\tau-1)\frac{
m_\rho^2}{q^2}\left(\frac{m_{\rho_2}^2}{q^2}\frac{m_{\rho_3}^2}{q^2}\right)
\end{eqnarray}
All these coincide with the prediction $1/q^{(2\tau-2)}$ from the
conformal symmetry \cite{polchinski} and the pertubative QCD
analysis \cite{brodsky}. From the above relations one can guess the
asymptotic behavior for generic states with radical number $n$ and
arbitrary twist $\tau$:
\begin{equation}
F^{(\tau)}_{nn}(q^2)\to (\tau-1)\frac{
m_\rho^2}{q^2}\left(\prod_{k=n}^{k=n+\tau-3}\frac{m_{\rho_k}^2}{q^2}\right).
\end{equation}
Or using $m_{\rho_n}^2=4n+4$ this can be simplified as:
\begin{equation}
F^{(\tau)}_{nn}(q^2)\to
(\tau-1)\frac{(n+\tau-2)!}{(n+1)!}\left(\frac{
m_\rho^2}{q^2}\right)^{\tau-1}.
\end{equation}
For $n=0$ this reduces to
\begin{equation}
F^{(\tau)}_{nn}(q^2)\to (\tau-1)!\left(\frac{
m_\rho^2}{q^2}\right)^{\tau-1}
\end{equation}
which has been derived analytically in Ref. \cite{brodsky4}.

\section{summary}
In this paper we employ the AdS/QCD models with linear Regge
trajectories and calculate the couplings of $g_{\rho HH}$ in this
typical holographic model, which is often referred as soft-wall
model or harmonic oscillator model. The calculated result of the
couplings $g_{\rho HH}$ shows that $g_{\rho HH}$ grows linearly with
the radical number $n$ and the twist of $H$, and thus are far from
being universal. A simple argument has been given for this behavior
based on quasi-classical picture of excited hadrons. As an
inference, the vector meson dominance is strongly violated, except
for the pion, of which the form factor is completely saturated by
the $\rho$ meson. the contributions of various $\rho$ mesons indeed
show a sign-alternating feature, which is supported by QCD analysis
\cite{shifman}. The asymptotic behavior of the form factor for
various states are studied and are in accordance with the
expectation from conformal symmetry.

\end{document}